**GPU-accelerated FREDopt package for simultaneous dose and LET$_d$ proton radiotherapy plan optimization via superiorization methods**

Damian Borys[1,2*], Jan Gajewski[2*], Tobias Becher[3,4,8], Yair Censor[7], Renata Kopeć[2], Marzena Rydygier[2], Angelo Schiavi[6], Tomasz Skóra[5], Anna Spaleniak[2], Niklas Wahl[3,4], Agnieszka Wochnik[2], Antoni Ruciński[2]

[1] Silesian University of Technology, Department of Systems Biology and Engineering, Gliwice, Poland
[2] Institute of Nuclear Physics PAN, Krakow, Poland
[3] Department of Medical Physics in Radiation Oncology, German Cancer Research Center (DKFZ), Heidelberg, Germany,
[4] Heidelberg Institute for Radiation Oncology (HIRO) and National Center for Radiation Research in Oncology (NCRO), Heidelberg, Germany,
[5] National Oncology Institute, National Research Institute, Krakow Branch, Krakow, Poland
[6] Dipartimento di Scienze di Base e Applicate per l'Ingegnera, La Sapienza Università di Roma, Rome, Italy
[7] Department of Mathematics, Faculty of Natural Sciences, University of Haifa, Haifa, Israel
[8] Department of Physics and Astronomy, Heidelberg University, Heidelberg, Germany

## Abstract

This study presents FREDopt, a newly developed GPU-accelerated open-source optimization software for simultaneous proton dose and dose-averaged LET (LET$_d$) optimization in IMPT treatment planning. FREDopt was implemented entirely in Python, leveraging CuPy for GPU acceleration and incorporating fast Monte Carlo (MC) simulations from the FRED code.

The treatment plan optimization workflow includes pre-optimization and optimization, the latter equipped with a novel superiorization of feasibility-seeking algorithms. Feasibility-seeking requires finding a point that satisfies prescribed constraints. Superiorization interlaces computational perturbations into iterative feasibility-seeking steps to steer them toward a superior feasible point, replacing the need for costly full-fledged constrained optimization.

The method was validated on two treatment plans of patients treated in a clinical proton therapy center, with dose and LET$_d$ distributions compared before and after reoptimization. Simultaneous dose and LET$_d$ optimization using FREDopt led to a substantial reduction of LET$_d$ and (dose)×(LET$_d$) in organs at risk (OARs) while preserving target dose conformity. Computational performance evaluation showed execution times of 14-50 minutes, depending on the algorithm and target volume size—satisfactory for clinical and research applications while enabling further development of the well-tested, documented open-source software.

*Corresponding Authors: D.Borys damian.borys@polsl.pl; J.Gajewski jan.gajewski@ifj.edu.pl

## Introduction

Inverse treatment planning methods in radiation therapy nowadays employ intensity modulated beam delivery techniques, such as intensity-modulated radiation therapy (IMRT) with X-rays or intensity-modulated proton therapy (IMPT), substantially improving treatment plan conformity and reducing radiation exposure of normal tissues, including organs at risk (OAR) neighboring the planning target volume (PTV). While the benefits of these techniques made them widely applied in the clinic, physical and biological uncertainties of proton treatment planning with intensity-modulation have been further investigated. Importantly, IMPT treatment planning approaches must additionally consider the relative biological effectiveness (RBE) of protons that, in clinical routine is commonly averaged to the value 1.1. In fact, the RBE is estimated to increase even up to 1.7 at the end of a proton beam range as a function of a dose-averaged linear energy transfer ($LET_d$), a physical parameter describing the radiation quality (Paganetti, 2014). Overall, IMPT has been exploited to reduce both the physical and biological uncertainties, i.e., increase treatment plan robustness, and decrease RBE in OAR.

More than a decade ago, empirical variable RBE models based on clonogenic cell survival data were mostly considered for proton treatment planning, but in vitro-based models turned out not to be appropriate for clinical predictions of radiation effects on normal and tumor tissues (A. McNamara et al., 2020). Hence, more focus was directed towards the development of treatment planning approaches based on $LET_d$, relying on its quasi-linear relationship with RBE. $LET_d$ as a physics quantity can be precisely calculated with Monte Carlo simulations, while measurement of LET or similar microscopic physics quantities is possible with state-of-the-art radiation detectors (Muñoz et al., 2024; Smith et al., 2021; Stasica et al., 2023). Soon it was demonstrated that proper optimization of IMPT plans enables reducing $LET_d$ in OAR without changing physical dose distribution (Giantsoudi et al., 2013; Grassberger et al., 2011). This was further motivated by the debate on the importance of $LET_d$ distributions for the prediction of post-treatment radiographic changes (McIntyre et al., 2023), which eventually led to a recent Heidelberg trial investigating risk-minimizing treatment planning concept, i.e., INDIGO (INDIvidualized, model-Guided Optimization of proton beam TP for LGG treatment) strategy for biologically weighted proton treatment planning incorporating variable RBE modeling, $LET_d$ optimization, and a normal tissue risk minimization modeling (Sallem et al., 2024). Currently, following the first paper by Unkelbach (Unkelbach et al., 2016), several IMPT treatment planning approaches exploit the product of (dose)×($LET_d$) as an optimization objective. These are investigated and implemented in research and clinical versions of treatment planning systems (TPS) (Deng et al., 2021; McIntyre et al., 2023).

The goal of this paper is to demonstrate the functionality of our newly, from-scratch developed and implemented, open-source treatment plan optimization software for simultaneous proton dose and $LET_d$ optimization, encapsulated in the FREDopt (Fast paRticle thErapy Dose optimizer) package (*FREDopt*, 2025). Our FREDopt package is based entirely on Python and exploits GPU acceleration in many execution steps, including Monte Carlo (MC) simulations performed in the FRED (Fast paRticle thErapy Dose evaluator) MC code. We provide a detailed explanation of how data is processed and how clinical treatment plans are reoptimized.

In the sequel of this work, we refer to the term "optimization" in the expression "treatment plan optimization" in the way used in medical physics and radiation oncology nomenclature, as a clinically satisfactory solution to the inverse radiation therapy treatment planning problem, and not specifically as solving a mathematical optimization problem. Similarly, we refer to the "simultaneous dose and $LET_d$ optimization", as solving the inverse radiation therapy treatment planning problem by simultaneously considering dose and $LET_d$, and not specifically considering dose and $LET_d$ when solving a mathematical optimization problem.

The algorithmic novelty of the work presented here is that instead of using a classical nonlinear constrained optimization algorithm to solve the inverse treatment planning problem, we implement two algorithmic approaches that rely on feasibility-seeking, see, e.g., (Bauschke & Borwein, 1996) and on the superiorization algorithm (SA) methodology, see, e.g., (Censor, 2023). We elaborate on these algorithmic approaches later in the paper. Specifically, we used here the AMS (Agmon, Motzkin, Schoenberg) feasibility-seeking approach (Agmon, 1954; Motzkin & Schoenberg, 1954), and the superiorization approach (Barkmann et al., 2023). We implemented them on GPU and applied them here, for the first time, to solve a simultaneous dose and $LET_d$ treatment planning optimization problem, showing benefits over the use of classical constrained optimization methods.

Feasibility-seeking methods handle the treatment planning problem by aiming to find a solution vector of intensities that will obey the upper and lower bounds on the physical property under consideration (e.g., dose) in the irradiated voxels, depending on the structure to which they belong. This is done without imposing an objective function according to which a particular feasible solution point should be preferred. This approach offers some advantages over classical optimization techniques to solve complex treatment planning problems by simplifying the formulation of the original inverse problem while maintaining solution quality. Moreover, the voxel-constrained formulation for feasibility seeking is very suited for GPU implementation due to the chosen algorithm performing projections in parallel.

When the planner wants to minimize an objective function with constraining it to boundary condition, it will result with model of constrained optimization. However, imposing an exogenous objective function over the upper and lower bounds constraints will result in superiorization approach. The superiorization methodology is designed to replace the need to apply classical optimization techniques and offers to find a „superior" feasible point from the constraints set. Superior means here that the point will be feasible for the constraints but, at the same time, have an equal or lower (not necessarily optimal) value of the objective function than a feasible point that has been obtained by feasibility-seeking alone – without superiorization. In radiotherapy planning systems, the SA offers a further benefit over feasibility only seeking and over full fledged constrained optimization techniques.

Finally, we show the capability of FREDopt to reoptimize clinical treatment plans of patients treated in the Cyclotron Centre Bronowice (CCB, Kraków, Poland) proton therapy facility, leading to comparable dose distributions in the tumor with superior $LET_d$ and (dose)×($LET_d$) distributions in normal tissues compared to the clinical plans. The performance of the FREDopt package calculations is reported demonstrating its potential applications.

**Materials and Methods**

## 2.1 Proton facility and patient data

The Cyclotron Centre Bronowice (CCB) in Kraków, is the only proton therapy facility in Poland, delivering a proton beams for radiotherapy purposes. The Proteus C-235 therapy system (IBA, Belgium) provides therapeutic proton beams to two rotational gantries equipped with dedicated scanning nozzles for IMPT and one eye treatment room. CCB has been operating since 2016, and has applied proton therapy for over 1,500 patients, including ocular and pediatric patients.

Two proton IMPT patients treated in CCB proton therapy center have been selected for treatment plan reoptimization in the FREDopt. Both patients had been diagnosed with a skull base chordoma and individual treatment plans have been prepared to deliver 74 Gy(RBE) of therapeutic dose in 37 fractions to the PTV region of volume 100 ml (Patient 1) or 368 ml (Patient 2). Each IMPT plan consisted of 4 fields irradiating the patients' head from 4 gantry angles.

## 2.2 The FRED MC code and the FREDtools package

FRED is a fast Monte Carlo (MC) code developed at Sapienza University of Rome, Italy (*FRED MC*, 2025; Schiavi et al., 2017). The code utilizes GPU acceleration, enabling the MC simulation of proton treatment plans to be completed in just a few minutes. It includes features for carbon ion tracking (De Simoni et al., 2022) and for the scoring of positron emitter production (K. McNamara et al., 2022). A separate branch of the FRED code implements electromagnetic interactions involving electrons and photons for conventional therapy simulations (Franciosini et al., 2023). The FRED package has been commissioned and experimentally validated for clinical proton beam models used at the CCB in Krakow, Poland (Gajewski et al., 2021) and other facilities (Gajewski et al., 2020) and has been applied for 4D treatment plan evaluation (Wochnik et al., 2024), study of the PET application for the proton range monitoring (Borys et al., 2022; Brzezinski et al., 2023), and patient-specific quality assurance (Komenda et al., 2025).

Basic per-voxel scorers available in FRED include dose-to-material, dose-to-water, and dose-averaged LET ($LET_d$). Per-voxel spectra of given quantities, e.g., LET, deposited energy, or track length, can be stored in vectorized images (Gajewski et al., 2024). The $LET_d$ is calculated according to the equation (Polster et al., 2015):

$$LET_d = \frac{\sum_{events} \frac{dE}{dx} dE}{\sum_{events} dE} \cdot \frac{1}{\rho} \qquad , \tag{1}$$

where $dE$ is the energy deposited by a particle in a single event of length $dx$ in a material of density $\rho$. The $\frac{dE}{dx}$ within each event is calculated as the mean electronic energy loss per unit path length corresponding to the average kinetic energy between values at the pre- and post-step points. In practice, the energy loss is precalculated before the particle tracking, once the particle type and material properties are known, based on the proton electronic stopping power table for water (Seltzer, 1993). The LET is scored in $\frac{MeV \cdot cm^2}{g}$ units, and the method is equivalent to the method 'C' proposed by (Cortés-Giraldo & Carabe, 2015). Averaging the LET with the deposited energy or dose as the weighting factor causes that, unlike, e.g., the dose, the $LET_d$ is not additive in a given point of the radiation field. However, the numerator and denominator of Equation (1) are additive. Therefore, FRED MC enables storing these quantities separately.

Each quantity scored in FRED can be stored as a cumulative 3D matrix in a MetaImage format and as an influence matrix in a sparse format, i.e., containing only the non-zero data to reduce the file and memory occupancy. Technical details about handling the influence matrix data structure can be found in FRED documentation (*FRED MC*, 2025). In general, an influence matrix provides a given quantity distribution per single primary, separately for each pencil beam. The implementation in FRED enables an influence matrix to contain any number of components. For instance, the dose influence matrix (also called Dij matrix) is a single-component sparse matrix describing the dose distribution per primary proton for each pencil beam. In turn, the $LET_d$ influence matrix contains two components, the numerator and denominator of the dose-averaged LET distribution, both provided per primary proton. Equation (1), describing the dose-averaged LET can be transformed in such a way that the denominator is the sum of the dose deposited by protons, which can be assumed to be approximately equal to the total deposited dose.

FREDtools is an open-source set of Python functions and classes designed for the manipulation and analysis of scalar or vector images (*FREDTools*, 2025). It leverages the SimpleITK framework (Lowekamp et al., 2013; Yaniv et al., 2018) and utilizes ITK objects, providing access to all the features of the Insight Toolkit (ITK) (Johnson et al., 2015; McCormick et al., 2014; Yoo et al., 2002). The primary methods were created for analyzing images in MetaImage format (*.mha or *.mhd), generated by various Monte Carlo frameworks, but they can also be applied to other formats, including complete DICOM files. FREDtools offers a range of image manipulation capabilities, such as resampling, performing affine transformations, mapping DICOM structures onto 3D image masks, and conducting image analyses like dose-volume histogram (DVH) analysis, multithreaded gamma index analysis, and Bragg peak analysis.

## 2.3 Feasibility-seeking and superiorization algorithms for treatment plan optimization

A review of the current arena of inverse radiation therapy treatment planning reveals that numerical optimization methods are commonly used to solve the mathematical problem of IMPT treatment planning. Clinical treatment goals, i.e., treatment dose prescribed to the tumor and dose limits to the organs at risk are translated into the non-linear constraints used in the multi-criteria optimization procedure. This approach to IMPT inverse problems gives rise to several scenarios amenable to the application of various established optimization algorithms. The so-called local optimization methods include (i) gradient-based methods, e.g., the nonlinear conjugate gradient (Jiang et al., 2022; Yu-Hong & Zexian, 2024) or limited-memory Broyden-Fletcher-Goldfarb-Shanno (BFGS) with bounds (Byrd et al., 1995; Yuan & Lu, 2011),

(ii) gradient-free methods like Nelder-Mead (Gao & Han, 2012) or Powell's conjugate direction method (Powell, 1964). Global methods include, e.g., genetic algorithms (Holland, 1992), and particle swarm optimization (Kennedy & Eberhart, 2002), etc. However, all these methods, especially the global ones, can be computationally challenging for the constrained nonlinear optimization approach to IMPT inverse treatment problems when the number of voxel constraints is substantial.

Most clinical and research TPS usually apply on-the-shelf optimization algorithm packages. For instance, the MatRad software (Wieser et al., 2017) exploits an interior point optimization (IPOPT) libraries (Wächter & Biegler, 2006) that implement interior-point filter line search algorithm (Wächter & Biegler, 2006). The commercial Eclipse (Varian Medical Systems, Switzerland) TPS used at the CCB proton therapy center exploits simultaneous spot optimization (SSO) based on the dose difference optimization (DDO) algorithm proposed by (Lomax, 1999), or the Conjugate Gradient (CG) algorithm (Xu et al., 2020). Another commercially available software RayStation (RaySearch, Sweden) TPS uses a gradient-based quasi-Newton approximation of the Hessian of the Lagrangian updated with BFGS method (Janson et al., 2024). OpenTPS, an open-source TPS for research in radiation therapy and proton therapy (Wuyckens et al., 2023), implements a variety of iterative solvers: quasi-Newton methods, LBFGS algorithms (Byrd et al., 1995), gradient-based methods or a beamlet-free algorithm (Pross et al., 2024). The JulianA package (Bellotti, 2024), developed with GPU support in Julia language at the Paul Scherrer Institute in Switzerland (Bellotti et al., 2023, 2024), implements a new spot weight optimization algorithm that minimises a scalar loss function using a gradient-based optimization algorithm.

The algorithmic approach adopted in our present work is different from all the above options. We use the superiorization methodology (SM) which was developed from the investigation of feasibility-seeking models of some important real-world problems such as image reconstruction from projections (Humphries et al., 2017) and radiation therapy treatment planning (Herman et al., 2012). Feasibility-seeking algorithms, mainly projection methods (Bauschke & Borwein, 1996) generate iterative sequences that converge to a point in the feasible set. Their main advantage is that they perform projections onto the individual sets whose intersection is the feasible set and not directly onto the feasible set and the underlying situation is that such projections onto the individual sets are more manageable. When one wishes to find feasible points with a reduced, not necessarily optimal, value of an imposed objective function then the SM comes into play.

The principle of the SM is to interlace into the iterates of a feasibility-seeking process perturbations that steer the iterates toward superior (meaning smaller or equal) objective function values without losing the overall convergence of the perturbed iterates to a feasible point. To this end "bounded perturbations" are used. How all this is done is rigorously described in earlier papers on the SM, consult, e.g., (Censor, 2023), for a recent review, read also (Herman, 2020).

A key feature of the SM is that it does not aim for a constrained optimal function value, but is content with settling for a feasible point with reduced objective function value – reduced in comparison to the objective function value of a feasible point that would be reached by the same feasibility-seeking algorithm without perturbations. This is sufficient for many applications, in particular, whenever the introduction of an objective function is only a secondary goal. Fulfillment of the constraints, in this context, is considered by the modeler of the real-world problem to be much more important[1]. Many papers on the SM are cited in (Censor, 2025) which is a website dedicated to superiorization and perturbation resilience of algorithms that contains a continuously updated bibliography on the subject. This Webpage is a source for the wealth of work done in this field to date, including a journal special issue (Gibali et al., 2020) dedicated to research of the SM. In radiotherapy treatment planning, superiorization was recently reported to achieve similar dosimetric performance to nonlinear constrained optimizers while ensuring smooth convergence (Barkmann et al., 2023). The work that we present here builds, from the algorithmic point of view only, upon that paper.

## 2.4 Implementation of simultaneous dose and LET$_d$ superiorization for treatment plan optimization in the FREDopt package

The FREDopt (Fast paRticle thErapy Dose optimizer) package (*FREDopt*, 2025) is an open-source Python software implementation allowing simultaneous optimization of physical dose and the product of physical dose and LET$_d$. The main part of the framework is entirely written in Python and when possible, executed on GPU exploiting Python CuPy libraries. The simultaneous optimization is effectively addressed by reoptimization of a dose-optimized treatment plan by reoptimizing the quantity (c)×(dose)×(LET$_d$), hereafter denoted as cDL, using a dedicated set of constraints specific for the cDL space. The included parameter c is a constant number equal to 0.04 μm/keV, following Unkelbach et al. (Unkelbach et al., 2016). The developed FREDopt package is capable of treatment plan optimization, including recalculation of beam positions, for predefined gantry angles, table rotations, and the isocenter. However, because this work aimed at demonstrating the feasibility of platform to time-efficient reoptimization of the existing clinical plans, the beamlets, gantry angles, table rotations, and the isocenter from the original clinical treatment plan are reused, and only the pencil beam fluence is reoptimized.

The implementation of the FREDopt treatment plan optimization framework is addressed in two subsequent stages, denoted as pre-optimization and optimization phase in Figure 1, and described in more detail in the following sections. Note that the pre-optimization and optimization expressions refer here to the stages of the treatment plan pre-optimization and optimization procedure. The word optimization does not specifically refer here to the optimization algorithm that is usually executed in the optimization phase - in this paper, instead, feasibility-seeking and superiorization algorithms are executed in the optimization phase.

The pre-optimization phase is preceded by the collection of the clinical data and information that is considered as the reference and starting point for treatment plan reoptimization. This phase includes data conversion from DICOM to the input files compatible with FRED MC and FREDopt. This phase also includes recalculating the treatment plan with FRED MC to obtain the $LET_d$ distributions needed to define treatment plan optimization constraints in the cDL space. The pre-optimization phase 1 covers the preparation and loading of all necessary data to the FREDopt package and the calculation of dose and $LET_d$ influence matrices in the FRED MC. Phase 2 is the actual treatment plan reoptimization phase that focuses on the execution of the feasibility-seeking and superiorization algorithms on GPU.

---

[1] Support for the reasoning of the SM may be borrowed from the American scientist and Noble-laureate Herbert Simon who was in favor of "satisficing" rather than "maximizing". Satisficing is a decision-making strategy that aims for a satisfactory or adequate result, rather than the optimal solution. This is because aiming for the optimal solution may necessitate needless expenditure of time, energy and resources. The term "satisfice" was coined by Herbert Simon (Simon, 1956), see also: https://en.wikipedia.org/wiki/Satisficing.

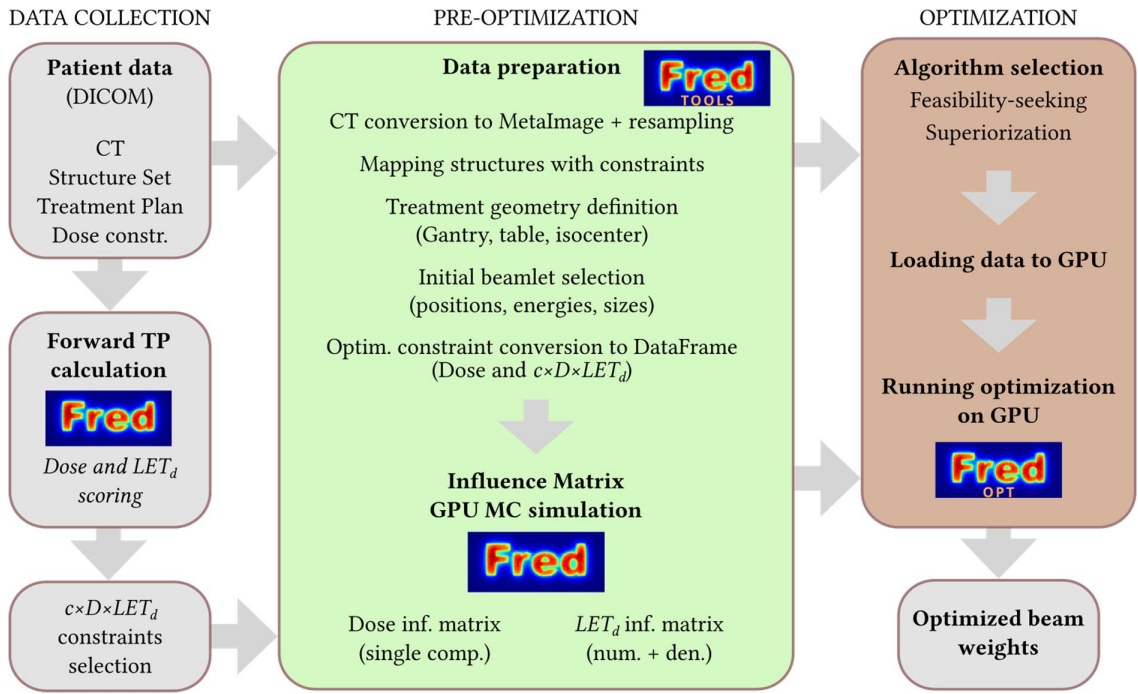

Figure 1. FREDopt general workflow.

## 2.4.1 Data collection

### Patient data

The input data to the MC calculation and treatment plan optimization process are anonymized and exported for a given patient from the clinical TPS Eclipse v. 16.1. The data in DICOM format include the patient's CT, structure set, and the clinical treatment plan. Additionally, clinical treatment plan optimization constraints and their weights are extracted to be reused in the treatment plan reoptimization. The PTV constraint implemented at CCB, which is more stringent than ICRU recommendations, mandates that at least 98% of the PTV receives 95% of the prescribed dose.

### Forward TP calculation and constraint selection

As a part of the preparation of the treatment plan reoptimization, the optimization constraints need to be additionally defined in the cDL space. This requires the forward calculation of the clinical TPS-generated treatment plan in the FRED MC code to obtain MC calculated dose, $LET_d$, and finally, the cDL distributions. These distributions are processed using the FREDtools package and dose-volume histogram (DVH), $LET_d$ volume histogram (LVH), and c×D×$LET_d$ volume histogram (cDLVH) are generated. In order to reduce the (c)×(dose)×($LET_d$) in the OARs, additional upper constraints equal to cDL30, i.e., the (c)×(dose)×($LET_d$) value received by 30% of the structure volume, were set, which should result in a reduction of the high $LET_d$ in the respective OAR.

## 2.4.2 Treatment plan pre-optimization phase

### FREDOpt input data preparation

Patient data is converted from DICOM format to the format compatible with FRED MC and the FREDOpt package. The CT is converted from dicom files to the Metalmage format, and resampled to the user-defined resolution, here 3x3x3mm$^3$. The treatment plan, describing pencil beam nominal energies and positions, irradiation geometry (fields, gantry and couch angles, isocenter position) is converted from DICOM to FRED-specific treatment plan data format applying the validated proton beam model of CCB Krakow proton therapy center (see Gajewski et al. (Gajewski et al., 2021)). Using the converted data and a CT calibration, specifying the conversion from HU CT values to material properties (composition, density, relative proton stopping power), the MC FRED simulation is performed utilizing the GPU acceleration, scoring the dose-to-water and $LET_d$ influence matrices.

## 2.4.3 Treatment plan optimization phase

The second step of the FREDopt workflow is the treatment plan optimization with the selected feasibility-seeking or superiorization algorithms. For feasibility-seeking, we have adapted and implemented the AMS algorithm, presented by Barkmann et al. (Barkmann et al., 2023), including simultaneous dose and (c)×(dose)×($LET_d$) optimization problem. In contrast to the original implementation that utilized only the dose influence matrix, both the dose and (c)×(dose)×($LET_d$) influence matrix were used additionally, along with the corresponding dose and cDL constraints. The pseudocode is presented in Figure 2.

```
function AMSrelaxDLET ( fluence, D, cDLET, U_D, L_D, U_cDLET, L_cDLET, λ, w ):

        # Dose

        for all ROI in ROIs do:                          # process each structure

                for all i ∈ ROI do:                      # process each voxel in ROI

                        if ⟨D_i, fluence⟩ > U_{D,i} :    # check upper bound if exists

                                fluence = fluence - λw_i $\frac{\langle D_i, \text{fluence}\rangle - U_{c,i}}{\|D_i\|_2^2}$ D_i

                        end if

                        if ⟨D_i, fluence⟩ < L_{D, i} :   # check lower bound if exists

                                fluence = fluence - λw_i $\frac{L_{cDLET} - \langle D_i, \text{fluence}\rangle}{\|D_i\|_2^2}$ D_i

                        end if

                end for

        end for

        # cDLET

        for all ROI in OARs do:                          # process each structure

                for all i ∈ ROI do:                      # process each voxel in ROI

                        if ⟨cDLET_i, fluence⟩ > U_{cDLET,i} :   # check upper bound if exists

                                fluence = fluence - λw_i $\frac{\langle \text{cDLET}_i, \text{fluence}\rangle - U_{c}\text{DLET}}{\|cDLET_i\|_2^2}$ cDLET_i

                        end if

                        if ⟨cDLET_i, fluence⟩ < L_{cDLET,i} :   # check lower bound if exists

                                fluence = fluence - λw_i $\frac{L_{cDLET,i} - \langle \text{cDLET}_i, \text{fluence}\rangle}{\|cDLET_i\|_2^2}$ cDLET_i

                        end if

                end for

        end for

        for all j ∈ beams do:                            # ensure fluence_j > 0

                if fluence_j < 0 :

                        fluence_j = 0

                end if

        end for

        return fluence

end
```

Figure 2. The pseudocode of the AMS algorithm, i.e., the feasibility-seeking algorithm for simultaneous dose and (c)×(dose)×(LET$_d$) optimization.

$U_{cDLET}$ and $L_{cDLET}$ input parameters are upper and lower constraints for all organs or defined for each point in the image space. $\lambda$ is a user-defined relaxation parameter and is limited to $0 < \lambda \leq 2$, and w is the weight (or priority) of each voxel, typically defined as a single weight for the whole structure (OAR or PTV) defined in the treatment plan.

The implementation of the SA uses the feasibility-seeking method as the main step of each iteration. However, it inserts the perturbations phase into it, which calculates the objective function value and its (negative) gradient. This additional step, performed between subsequent feasibility-seeking iterations, yields locally a reduction of the function value due to the move in the negative gradient direction. By integrating such perturbations into the iterative steps of the feasibility-seeking algorithm, SA is capable of improving solution quality by reaching a feasible point with a lower (not necessarily optimal) objective function value. This replacement of the quest from accurate constrained optimization by feasibility with lower function value causes computational efficiency without compromising the quality of the solution in terms of the treatment planning. The pseudocode is shown in Figure 3.

```
function SuperiorDLET ( fluence, cDLET, U_cDLET, L_cDLET, λ, w, α, η ):

        k = 0; s= -1;

        fluence^k = fluence

        while k < iterMax do:                      # or any other stopping rule

                t = 0                              # start perturbation step

                fluence^{k,t} = fluence^k

                while t < reductionsN do:          # define number of reductions

                        loop = True

                        while loop do :

                                s = s + 1

                                β = α^s             # step size for gradient updates

                                fluence_new = fluence^{k,t} - β·grad_cost( fluence^{k,t} )

                                # perform reduction with gradient function

                                if cost( fluence_new ) < cost( fluence^{k,t} ) :

                                        t = t + 1

                                        fluence^{k,t} = fluence_new

                                        loop = False

                                end if

                        end while

                end while

                w_k = η^k w                        # reduce the weight in each iteration

                # perform feasibility seeking step

                fluence^{k+1} = AMSrelaxDLET( fluence^k, D, cDLET, U_D, L_D, U_cDLET,
                L_cDLET, λ, w_k )

                k = k + 1

        end while

        return fluence^k

end
```

Figure 3. The pseudocode of SA (superiorization algorithm) for simultaneous
(c)×(dose)×(LET_d) optimization.

Where the parameter α is a fixed user-defined constant $0 < α < 1$ that controls β, and β is a step size for the negative gradient perturbations defined by $α^s$. In each superiorization iteration, the weights for each organ are reduced by the factor $η^k$ where $0 < η < 1$.

This method requires objective function calculation. As we aim for simultaneous dose and (c)×(dose)×(LET_d) optimization, we include both factors in our function, which is described by Equation (2).

$$\chi(x)^2 = \sum_{i \in PTV} w_i(\widehat{D}_i - D(x)_i)^2 + \sum_{i \in OAR} w_i(\widehat{cDL}_i - cDL(x)_i)^2 \quad ,$$
(2)

where $\chi(x)^2$ is the cost function with the argument $x$ - the fluence vector, $D(x)_i$ and $cDL(x)_i$ are the dose and (c)×(dose)×(LET$_d$) values in the voxel $i$, that belong to PTV or OAR respectively, while $\widehat{D}_i$ and $\widehat{cDL}_i$ are constraints defined by the user or calculated from forward TP calculation step; $w_i$ are weights for each organ. Note that the constraints used for optimization were obtained from the initial clinical plan recalculation exploiting dose and cDL volume histogram data. The DVH constraints are used for the target voxels which are defined as a PTV, and cDL-VH constraints are used for all OARs.

The optimization constraints and weights are extracted, reused, and adjusted as needed. For example, the cDL constraints are modified in order to reduce the cDL values in the organs at risk. This adjustment can be applied to any region of interest without significantly affecting the treatment planning time, as it mainly depends on the number of iterations required.

The FREDopt code has been implemented to be used with graphics accelerators available to the general public, similarly to the FRED MC code. Our calculations were performed on the NVIDIA P100 card with 16 GB of RAM, installed in a server located at the Silesian University of Technology. However, any CUDA-enabled card can be used, and the limitation is the memory capacity - most plans were able to fit within the 12 GB of RAM on the graphics card, and some require up to 16 GB.

All calculations were performed using the resolution of 3x3x3 mm$^3$ mainly due to energy step and scanning grid resolution but also due to memory limitations. The final dose and LET$_d$ grid correspond the resolution of the CT used in MC calculation.

## 2.5 Evaluation of the treatment plans

To evaluate the reoptimized plans, we compared the spatial dose and (c)×(dose)×(LET$_d$) distributions, calculating the difference between the dose only and simultaneous (c)×(dose)×(LET$_d$) re-optimized treatment plans outputs, that are obtained after the second stage of the workflow. The visual comparison is done on the basis of the corresponding Dose Volume Histograms (DVH) and (c)×(dose)×(LET$_d$) Volume Histograms (cDLVH). The reoptimized plans are compared after using the AMS and its superiorization algorithm to unveil the differences in those approaches. The performance of both algorithms is also evaluated by showing calculation time results of the treatment plan optimization procedure for each execution stage.

A robustness analysis has been performed for the reoptimized plans. The analysis follows the protocol used at the CCB and involves recalculating nine plans: the initial setup, variations with geometrical shifts of +/- 2 mm in each direction, and CT density shifts of +/- 3.5%. The robustness of the dose and (c)×(dose)×(LETd) for the reoptimized plans was compared with the initial TPS plan for the CTV structure.

# Results

Forward calculation of the clinical treatment plan in the FRED MC code allowed for the acquisition of dose and cDL distributions, which were used to calculate (c)×(dose)×(LET$_d$) constraints needed for the second phase of the treatment plan reoptimization procedure. The statistics for PTV and selected organs at risk are presented in Table 1, along with the calculated cDL30 constraints.

Table 1. OAR statistics and constraints for two patient examples.

| | | statistics | | | | constraints |
|---|---|---|---|---|---|---|
| | structure | Volume [cm3] | cDL min [Gy] | cDL mean [Gy] | cDL max [Gy] | cDL30 [Gy] |
| Patient 1 | Brainstem | 29.6 | 0.04 | 2.82 | 10.05 | 3.1 |
| | Hippocampus Left | 5.2 | 0.34 | 5.69 | 12.63 | 8.4 |
| Patient 2 | Brainstem | 46.0 | 0.09 | 3.36 | 10.17 | 4.0 |
| | Optic Nerve Left | 3.6 | 0.66 | 4.60 | 8.98 | 6.0 |

Figures 4 and 5 present the spatial distributions of dose and (c)×(dose)×(LET$_d$) before and after treatment plan reoptimization, together with the relative differences between them, for patients 1 and 2, respectively. In both cases, critical structures and PTVs are highlighted. The final comparison is presented in the figures in the form of volume histograms, where the solid lines show the status before and the dashed and dotted lines the status after treatment plan reoptimization with feasibility seeking and superiorization algorithms, respectively.

In both presented cases, the PTV is located near the brainstem critical structure, where the treatment plan reoptimization in cDL space aimed to lower LET$_d$ distribution. The reduction of the high cDL values at the borders of the brainstem structure (marked in blue contour) can be observed for the example patient cases. This is even more clearly visible in the cDLVHs, where the maximum values of cDL were reduced due to the treatment plan reoptimization procedure. The difference distributions show a reduction of dose and cDL of about 20-30%. The dose and cDL reduction were also observed in other structures that are located in the proximity of the tumor volume (left hippocampus in Patient 1 and left optic nerve in Patient 2), however, they were not visible on the distributions as they are located on different slices.

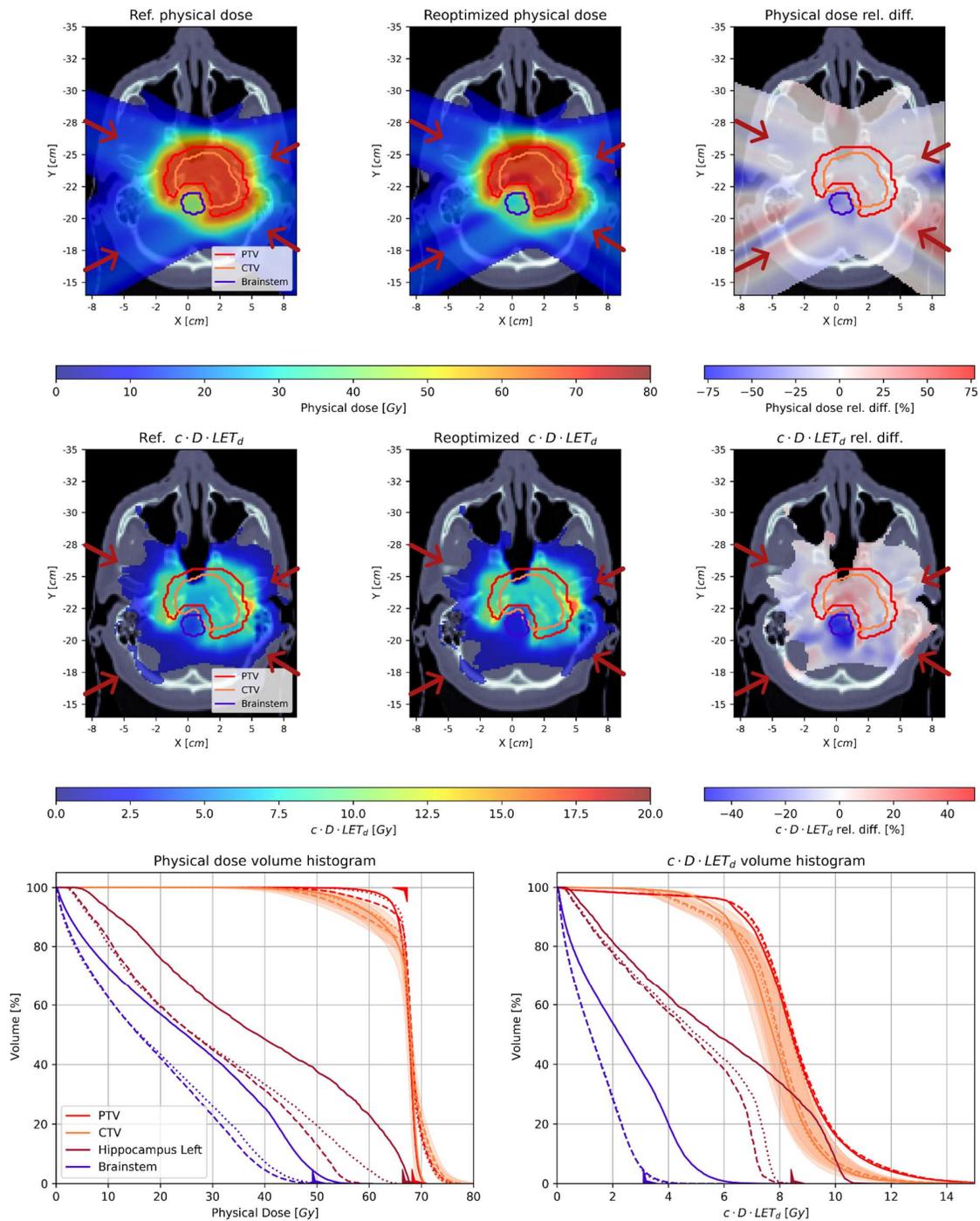

Figure 4. Results for patient 1. Upper row: physical dose distribution for (left) the initial plan, (middle) the reoptimized plan using the SA, and (right) the relative difference between them. Middle row: cDL distribution for (left) the initial plan, (middle) the reoptimized plan using the SA, and (right) the relative difference between them. Red arrows denote the directions of the four fields. Bottom row (left) DVH and (right) cDLVH, for the PTV (red), CTV (orange), and two OARs: the left hippocampus (brown) and the brainstem (blue). Solid, dashed, and dotted lines

denote the histograms for the initial, reoptimized using the AMS algorithm, and reoptimized using the SA plans, respectively. The orange shadows represent the robustness analysis for the CTV structure.

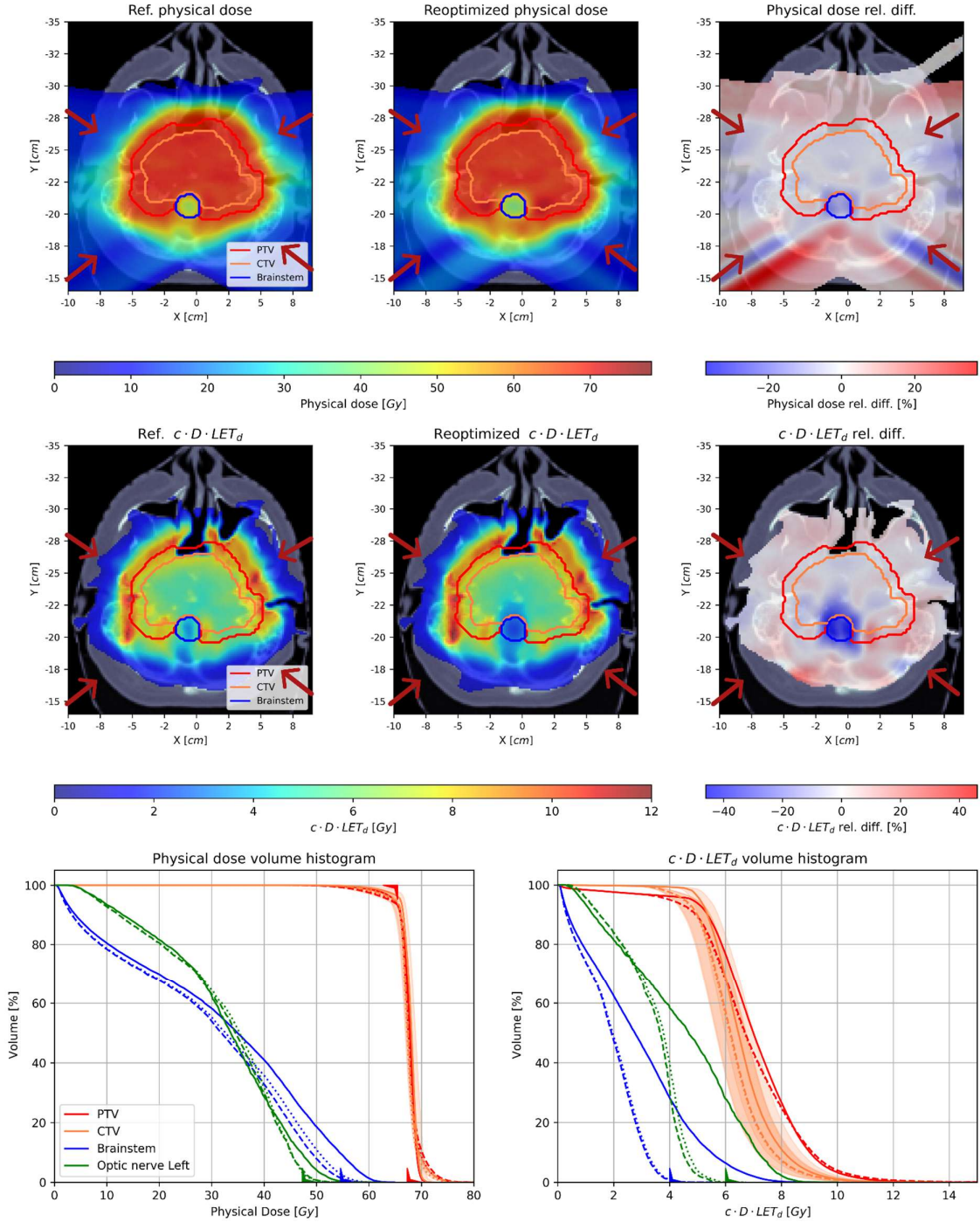

Figure 5. Results for patient 2. Upper row: physical dose distribution for (left) the initial plan, (middle) the reoptimized plan using the SA, and (right) the relative difference between them. Middle row: cDL distribution for (left) the initial plan, (middle) the reoptimized plan using the SA, and (right) the relative difference between them. Red arrows denote the directions of the four fields. Bottom row (left) DVH and (right) cDLVH, for the PTV (red), CTV (orange), and two OARs: the left optic nerve (green) and the brainstem (blue). Solid, dashed, and dotted lines denote the histograms for the initial, reoptimized using the AMS algorithm, and reoptimized using the SA plans, respectively. The orange shadows represent the robustness analysis for the CTV structure.

The entire workflow, as described above, can be divided into distinct stages with varying execution times. Here, we report the time for: (i) data preparation, (ii) influence matrix GPU MC simulation, (iii) loading and copying dose and $LET_d$ influence matrices to GPU memory, and (iv) execution of AMS or superiorization algorithms.

Table 2. Consecutive steps execution times in seconds for the two example patients, reoptimized with the AMS or superiorization (SA) algorithm with 200 iterations. The execution time for the following phases is reported: (i) data preparation, (ii) influence matrix GPU MC simulation, (iii) loading and copying dose and $LET_d$ influence matrices to GPU memory, and (iv) execution of AMS or superiorization algorithms.

| | | Processing step time [s] | | | | Total time |
|---|---|---|---|---|---|---|
| | algorithm | (i) | (ii) | (iii) | (iv) | [s] |
| **Patient 1** | **AMS** | 17 | 522 | 83 | 208 | 830 |
| | **SA** | | | 93 | 2091 | 2723 |
| **Patient 2** | **AMS** | 63 | 1362 | 104 | 143 | 1672 |
| | **SA** | | | 114 | 1508 | 3047 |

## Discussion

In this work, we demonstrated the functionality of FREDopt, a new open-source software package designed to solve the inverse IMPT treatment planning problem using feasibility-seeking and superiorization algorithms. Our results show that the GPU-implemented FREDopt package, supported by fast FRED MC simulations, can improve $LET_d$ distributions in OARs without compromising target dose through simultaneous reoptimization of clinical proton treatment plans. A substantial reduction in (c)×(dose)×($LET_d$) in OARs was achieved for two IMPT plans used for patient treatments at the CCB Krakow proton therapy center. The treatment plan reoptimization time with the feasibility-seeking algorithm (designated AMS) for treatment plans was 14 and 28 minutes for 100 ml and 368 ml PTVs, including the MC-based dose and $LET_d$ influence matrices calculation.

Note, that the dose, and especially, $LET_d$ influence matrices in the full representation in memory (non-sparse matrix) can be extremely large. The size of the matrix results from the number of beamlets and the number of voxels in the image space. For the $LET_d$ influence matrix, the numerator and denominator are stored, and then consequently loaded separately. The full-size representation of these arrays exceeds the size of standard PC memory and requires large computing units with 256 GB of RAM or more, and a high-end GPU accelerator, which would limit the applicability of the FREDopt. Obviously, this would not allow calculations to be performed with a mid-range GPU accelerator, typically equipped today with 12 GB of memory, or even a server GPU unit with memory up to 24 GB. Therefore these arrays are stored as sparse matrices. Despite this, for some patients with relatively large targets/structures, the GPU having at least 16 GB of RAM was required to store the sparse matrix. Management of those matrices can also be a time-intensive task, which can be seen in the execution times shown in Table 2. Especially for the case where the low number of iterations of the algorithm is sufficient to obtain satisfactory results, this step is the most time consuming within the whole procedure. Presented times were obtained for a computational server equipped with two CPUs with 10 cores each and access to fast disc space. The computation times of the two algorithms, seen in Table 2, differ in favor of the feasibility-seeking algorithm, which is due to the fact that superiorization iteratively uses feasibility-seeking, interleaving it with the perturbation phase. The number of iterations of the superiorization algorithm determines how much longer the entire algorithm will take to execute, for the same number of feasibility-seeking iterations. However, with the longer calculation time, SA allows for a slightly more conformal dose distribution and comparable cDL benefit, with respect to the feasibility-seeking algorithm (see Figures 4 and 5).

Analysing the results presented in Figures 4 and 5 we observed differences of approximately 10% after re-optimization, which manifest as hotspots in the dose distribution within the PTV. This leads to a slight deterioration in the dose distribution in the PTV, as indicated by worse dose homogeneity in the DVH. However, this comes at the advantage of a significant improvement in the LET distribution in critical areas, such as the Brainstem and Hippocampus, as shown in Figure 4.

The difference observed in Figure 4 of the DVH for the Left Hippocampus between the two algorithms can be seen in other organs as well, indicating a behavior that is specific to each patient. In this particular case, the difference arises from the proximity of the Hippocampus to the PTV, which is asymmetric, causing a shift in mass towards the Hippocampus. However, in most instances, we do not observe significant changes in the results obtained from both algorithms.

The example shown in Fig. 5 illustrates small dose hotspots within and around the PTV, particularly near the Brainstem. Additionally, there is a noticeable reduction in the dose at the edge of the Brainstem area. This reduction is even more pronounced in the cDLET distribution, where the decrease in the Brainstem is significant. Emerging hotspots in the dose distribution are also evident in the DVH, indicating that a small percentage of voxels with increased dose for the PTV in the re-optimized plan. The newly re-optimized plan, i.e., the high dose tail in the PTV and CTV, is clinically acceptable and justified especially considering the significant reduction in cDL for OARs, particularly in the Brainstem and Optic Nerve areas. The re-optimized plans were reviewed in collaboration with the medical doctor co-author, who confirmed their clinical relevance.

A separate issue is related to the precision of the $LET_d$ calculation in MC codes in general, and FRED MC in particular. The accuracy of single particle LET and the average $LET_d$ calculation in FRED MC has been recently validated experimentally against Timepix measurement results for IMPT treatment plans (Stasica-Dudek et al., 2025). The work reveals the relative difference between the calculated and measured $LET_d$ below 5%. The FRED MC calculation of LET has also been benchmarked against other general-purpose MC, namely Geant4 and FLUKA, and described in a publication that is in the review process.

The GPU-accelerated FRED MC code has been in development for a decade, and its extension with FREDopt for proton therapy treatment planning marks a step toward full TPS functionality. Simultaneous treatment plan optimization utilizing dose and $LET_d$ has been explored by several research groups, as reviewed by Deng et al. (Deng et al., 2021) and McIntyre et al. (McIntyre et al., 2023). A key advantage of FREDopt is its Python-based architecture, which leverages CuPy libraries to enable GPU acceleration at multiple stages of the treatment plan optimization process. Moreover, the FREDopt implementation is divided into pre-optimization and treatment plan optimization steps, with the latter utilizing feasibility-seeking and superiorization algorithms that have not been previously applied to simultaneous treatment plan optimization utilizing dose and $LET_d$. These algorithms, from scratch implemented in Python, demonstrate that computationally demanding inverse treatment planning tasks can be efficiently executed within a scripting language, and they can be effectively iterated on GPUs, with modern 12–16 GB cards capable of loading the influence matrix in sparse format.

The results presented here indicate that while the calculation times are reasonable, they remain lengthy, particularly for clinical applications and robustness optimization. This version of the FREDopt and FREDtools allows us to achieve the times reported in the paper. However, it is still a work in progress, and there are opportunities to improve efficiency in some steps. To reduce calculation times, we can adjust the grid size, which is currently set at 3x3x3 mm³. Opting for a coarser grid will shorten the FREDMC simulation and accelerate the loading of the Dij and LETij matrices. In a more recent version of the FREDtools, an optimized method for storing the Dij and LETij matrices has been introduced, which should enhance the speed of step (iii). Additionally, utilizing newer and more powerful GPU accelerators or increasing the number of GPUs will further decrease MC simulation times, as the FRED MC package supports multiple GPU usage. In summary, with a few adjustments, we can significantly reduce total processing time, bringing us closer to the clinical implementation of the FREDopt package.

The selected resolution of 3x3x3 mm³ is primarily due to the GPU memory limitations associated with the optimization process. While our code can process any CT resolution, using higher-resolution images would generate much larger Dij and LETij matrices. This would necessitate a GPU accelerator with additional RAM, which could also increase computation time. However, the selected resolution is adequate for plan optimization purposes. To our knowledge, clinical systems for plan optimization also utilize downsampled CT images, and the resolution selected in our manuscript aligns with the standards typically used in clinical settings. Additionally, considering the energy step and scanning grid, which are both approximately 3 mm, the use of this resolution is justified.

A key trend in IMPT treatment planning, as adopted by the vendors of state-of-the-art TPSs, is the use of the $LET_d$ as a physical surrogate for the RBE, allowing modulation of $LET_d$ distributions in the target and OARs without altering the target physical dose. While the reduction in (c)×(dose)×($LET_d$) in selected OARs was successful for both investigated algorithms presented in this manuscript, the clinical and medical physics question about the robustness of the $LET_d$-optimized treatment plans (see Fig. 4 and 5), or more generally, treatment planning protocol or guideline for simultaneous dose and $LET_d$ optimization, remains open for discussion. Strategies like INDIGO (Sallem et al., 2024), particularly when implemented in the clinical trial, have the potential to shed light on the importance of selected parameters, including $LET_d$. However, the importance of RBE and $LET_d$ distributions, along with the proton beam robustness to range uncertainties, including the selection of beamlets and field directions, leave the space of treatment plan optimization parameters relatively large and prone to misinterpretations when studying only the selected parameters. The free parameter space can be even larger when considering more advanced arc delivery or upright immobilization techniques. The lack of consensus within the proton therapy community and unresolved challenges in biologically weighted treatment planning, particularly in ion therapy, highlight the need for further development of innovative planning approaches, including efficient and adaptable research TPS. While projects like MatRad and OpenTPS address various needs and serve multiple users, we believe our Python-based open-source FREDopt package, featuring novel algorithms and GPU acceleration, is a valuable tool for researchers and clinical medical physicists to support investigation of new strategies for biologically weighted proton therapy treatment planning. Making FREDopt available to the community paves the way for broader adoption and collaborative improvements of proton therapy treatment planning.

## Conclusions

We have demonstrated the functionality of FREDopt, an open-source, GPU-accelerated software package for simultaneous proton dose and $LET_d$ optimization. By integrating feasibility-seeking and superiorization algorithms, FREDopt provides an alternative to conventional nonlinear constrained optimization methods, offering a computationally efficient approach to treatment plan reoptimization.

The Python-based architecture, combined with GPU acceleration, enables rapid execution, making the software suitable for clinical applications. The superiorization algorithm and the AMS feasibility-seeking method effectively reduce $LET_d$ in OARs while preserving tumor dose conformity. Clinical validation on IMPT plans from CCB Kraków demonstrated improved $LET_d$ distributions compared to standard treatment plans, supporting the potential clinical relevance of this approach.

With the ongoing open-access strategy of FRED MC, making FREDopt available to the research and clinical community fosters broader adoption, validation, and collaborative development. Future work will focus on integrating micro- and nanoscale biologically weighted treatment planning methods for proton and ion therapy, while further enhancing optimization speed and robustness.

## Acknowledgments

This research was funded by the Polish Ministry of Education and Science under grant No. NdS/544748/2021/2021. The work of Yair Censor, Tobias Becher and Niklas Wahl is supported by the Cooperation Program in Cancer Research of the German Cancer Research Center (DKFZ) and Israel's Ministry of Innovation, Science and Technology (MOST) and by the U.S. National Institutes of Health Grant Number R01CA266467.
The authors are indebted to Jan Unkelbach for his advice.

## Author Contribution

DBo: Implementation of the FREDopt package, data analysis, interpretation, visualization, drafting of the manuscript
JGa: Data collection, implementation of the FREDtools and FREDopt packages, data analysis and interpretation, visualization, and drafting of the manuscript
TBe: Consulting of algorithms implementation
YCe: Consulting of algorithms implementation and drafting parts of the manuscript
RKo: Consulting in medical physics
MRy: Consulting in medical physics and data collection
ASc: Implementation of influence matrix calculations
TSk: Radiation oncology consulting and clinical data collection
NWa: Consulting of algorithms implementation
AWo: Data analysis
ASp: Data collection
ARu: Conceptualization and design of the study, drafting the manuscript, acquisition of funding
All authors reviewed and accepted the final version of the manuscript.